\documentclass{elsart}
\usepackage{graphics,epsfig,amssymb}

\begin{document}

\begin{frontmatter}

\title{Dip in UHECR spectrum as signature of proton interaction with CMB}

\author[LNGS]{V. Berezinsky},
%\ead{veniamin.berezinsky@lngs.infn.it}
\author[DESY]{A. Z. Gazizov} and
%\ead{gazizov@ifh.de}
\author[INR]{S. I. Grigorieva}
%\ead{grigorieva@inr.npd.ac.ru}

\address[LNGS]{INFN, Laboratori Nazionali del Gran Sasso,
I-67010 Assergi (AQ) Italy}
\address[DESY]{DESY Zeuthen, Platanenallee 6, D-157 Zeuthen, Germany}
\address[INR]{Institute for Nuclear Research, 60th October Revolution prospect
7A,\\ 117312 Moscow, Russia}

\begin{abstract}
Ultrahigh energy (UHE) extragalactic protons propagating through
cosmic microwave radiation (CMB) acquire the spectrum features in
the form of the dip, bump (pile-up protons) and the
Greisen-Zatsepin-Kuzmin (GZK) cutoff. We have performed the
analysis of these features in terms of the modification factor.
This analysis is weakly model-dependent, especially in case of the
dip. The energy shape of the dip is confirmed by the Akeno-AGASA
data with $\chi^2=19.06$ for $\mbox{d.o.f.} =17$ with two free
parameters used for comparison. The agreement with HiRes data is
also very good. This is the strong evidence that UHE cosmic rays
observed at energies $1\times 10^{18}$~eV -- $4\times 10^{19}$~eV
are extragalactic protons propagating through CMB. The dip is also
present in case of diffusive propagation in magnetic field.

\end{abstract}

\end{frontmatter}

\section{Introduction}
\label{intr}
The nature of signal carriers of UHECR is not yet
established. The most natural primary particles are extragalactic
protons. Due to interaction with the CMB radiation the UHE protons
from extragalactic sources are predicted to have a sharp
steepening of energy spectrum, so called GZK cutoff \cite{GZK}.

There are two other signatures of extragalactic protons in the
spectrum: dip and bump \cite{HS85,BG88,YT93,Stanev00}.
The dip is produced due to
$p+\gamma_{\rm CMB} \to p+e^++e^-$ interaction at energy centered by
$E \approx 8\times 10^{18}$~eV. The bump is produced by pile-up protons
which loose energy in the GZK cutoff. As was demonstrated in
\cite{BG88}, see also \cite{Stanev00}, the
bump is clearly seen from a single source at large redshift $z$, but
it practically disappears in the diffuse spectrum, because individual
peaks are located at different energies.

As will be discussed in this paper, the dip is a reliable feature in the
UHE proton spectrum (see also \cite{BGG} - \cite{BGG2}).
Being relatively faint feature, it is however clearly seen in the
spectra observed by AGASA, Fly's Eye,
HiRes and Yakutsk arrays (see \cite{expdata} and \cite{NW} for the data).
We argue here that it can be
considered as the confirmed signature of interaction of extragalactic
UHE protons with CMB.
\begin{figure}[ht]
  \begin{center}
    \includegraphics[width=8.0cm]{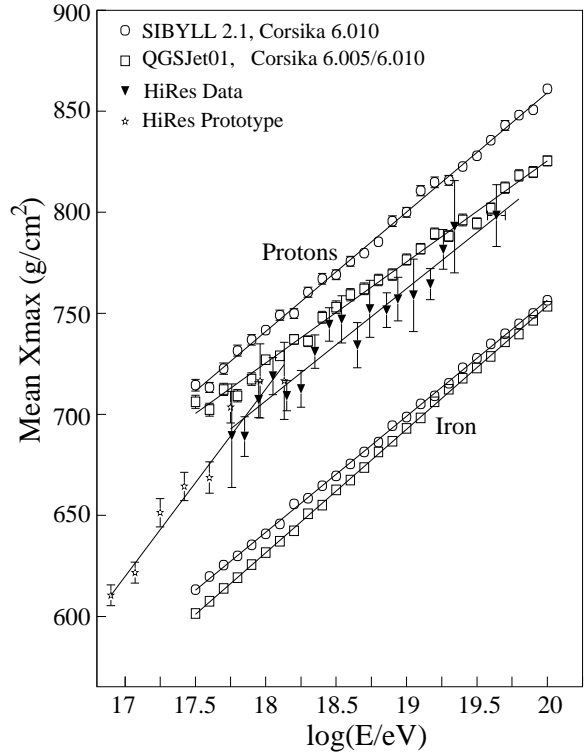}
\end{center}
  \caption{The HiRes data \cite{mass-Hires} on mass
composition. The measured atmospheric depths of EAS maximum $X_{\rm max}$ at
$E\geq 1\times 10^{18}$~eV (triangles) are in a good agreement with
QGSJet-Corsika prediction for protons.}
  \label{mass-comp}
\end{figure}

The measurement of the
atmospheric height
of EAS maximum, $X_{\rm max}$, in the HiRes experiment
(see Fig.\ref{mass-comp}) gives another evidence of the proton
composition of UHECR at $E \geq 1\times 10^{18}$~eV.
Yakutsk \cite{Glushkov00}
and HiRes-Mia \cite{Hi-Mia} data also favour
the proton composition at $E\geq 1\times 10^{18}$~eV,
though some other methods of mass measurements indicate
the mixed chemical composition \cite{Watson04}.

At what energy the extragalactic component sets in?

According to the KASCADE data \cite{Kampert01}, the spectrum of
galactic protons has a steepening at $E\approx 2.5\times
10^{15}$~eV (the first knee), helium nuclei - at $E \approx
6\times 10^{15}$~eV, and carbon nuclei - at $E \approx 1.5\times
10^{16}$~eV. It confirms the rigidity-dependent confinement with
critical rigidity $R_c=E_c/Z \approx 3\times 10^{15}$~eV. Then
CR galactic iron nuclei are expected to have the critical energy of
confinement at $E_c \sim 1\times 10^{17}$~eV, and extragalactic
protons can naturally dominate at $E \geq 1\times 10^{18}$~eV.
This energy is close to the position of the {\em second knee} (Akeno -
$6\times 10^{17}$~eV, Fly's Eye - $4\times 10^{17}$~eV, HiRes -
$7\times 10^{17}$~eV and Yakutsk - $8\times 10^{17}$~eV).  The
detailed analysis of transition from galactic to extragalactic
component of cosmic rays is given in \cite{BGH}. It favours the transition
at $E \sim 1\times 10^{18}$~eV. The model of galactic cosmic rays
developed by Biermann et al \cite{Bier} also predicts the second
knee as the ''end'' of galactic cosmic rays  due to
rigidity bending in wind-shell around SN, produced by Wolf-Rayet
stars. The extragalactic component becomes the dominant one at
energy $E \sim 1\times 10^{18}$~eV (see Fig.1 in \cite{Bier}).

Below we shall analyze the features in UHE proton spectrum using
basically two assumptions: the uniform distribution of the sources in the
universe and the power-law generation spectrum. We shall discuss how
large-scale inhomogeneities in source distribution affect the shape
of the features. We do not consider the possible speculations, such as
cosmological evolution of sources.
In contrast to our earlier works \cite{BGG,BGG1,BGH}, we do not use
here the model-dependent complex generation spectrum: the modification
factor method allows us to use more general power-law spectrum with an
arbitrary $\gamma_g$.
\vspace{-5mm}
\section{Bump in the diffuse spectrum}
\label{bump}
\vspace*{-5mm}
The analysis of the bump and dip is convenient to perform in terms
of {\em modification factor} \cite{BG88}.

The modification factor is defined as a ratio of the spectrum
$J_p(E)$, with all energy losses taken into account, to unmodified
spectrum $J_p^{\rm unm}$, where only adiabatic energy losses (red
shift) are included,
\begin{equation}
\eta(E)=J_p(E)/J_p^{\rm unm}(E).
\label{modif}
\end{equation}
For the power-law generation spectrum $\propto E_g^{-\gamma_g}$
from the sources without cosmological evolution one obtains the
unmodified spectrum as
\begin{equation}
J_p^{\rm unm}(E)=\frac{c}{4\pi}(\gamma_g -2) {\mathcal L}_0
E^{-\gamma_g} \int_0^{z_{\rm max}}dz
\frac{dt}{dz}(1+z)^{-\gamma_g+1}, \label{unm}
\end{equation}
where the observed energy $E$ and emissivity ${\mathcal L}_0$ are
measured in GeV and GeV/Mpc$^3$yr, respectively.  The connection
between $dt$ and $dz$ is given by usual cosmological expression (see e.g.
\cite{BGG1}). The flux $J_p(E)$ is calculated as in \cite{BGG}  with
all energy losses included.
\begin{figure}[ht]
\begin{center}
\epsfig{figure=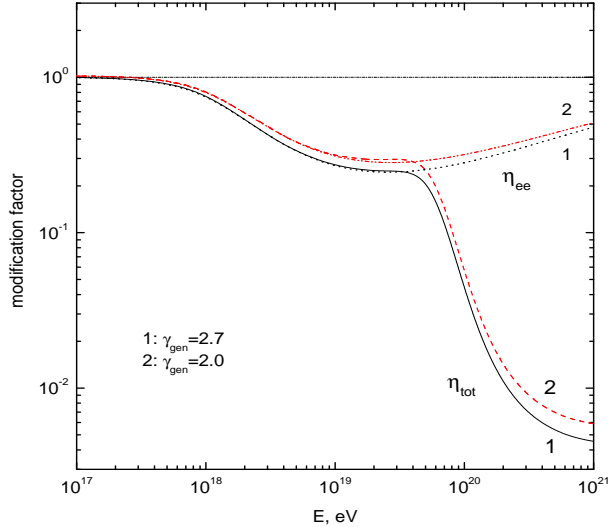,height=7cm,width=8cm}
\end{center}
%\vspace{-10mm}
\caption{
Modification factor for the power-law generation spectra with
$\gamma_g$ in the range 2.0 - 2.7. Curve $\eta=1$ corresponds to adiabatic
energy losses, curves $\eta_{ee}$ - to adiabatic and
pair production energy losses and curves $\eta_{\rm tot}$ ~-~ to all
energy losses.
}
\label{mfactor}
\end{figure}

In Fig.~\ref{mfactor} the modification factor is shown as a function
of energy for two spectrum indices $\gamma_g=2.0$ and $\gamma_g=2.7$.
They do not differ much from each other because both numerator and
denominator in Eq.~(\ref{modif}) include factor $E^{-\gamma_g}$.
Let us discuss first the bump. We see no indication of the bump
in Fig.~\ref{mfactor} at merging of $\eta_{ee}(E)$ and $\eta_{\rm
tot}(E)$ curves, where it should be located. The absence of the
bump in the {\em diffuse spectrum} can be easily understood. The
bumps are clearly seen in the spectra of the single remote sources
\cite{BG88}. These bumps, located at different energies, produce
a flat feature, when they are summed up in the diffuse spectrum.
This effect can be illustrated by Fig.~5 from
Ref.~\cite{BG88}. The diffuse flux there is calculated
in the model where sources are distributed uniformly in the sphere
of radius $R_{\rm max}$ (or $z_{\rm max}$). When $z_{\rm max}$
are  small (between 0.01 and 0.1) the
bumps are seen in the diffuse spectra. When radius of the sphere
becomes larger, the bumps merge producing the flat feature in the
spectrum.  If the diffuse spectrum is
plotted as $E^3J_p(E)$ this flat feature looks like a
pseudo-bump.
\vspace*{-5mm}
\section{Dip as a signature of the proton interaction with CMB.}
\vspace*{-5mm}
The {\em dip} is more reliable signature of interaction of protons
with CMB than GZK feature. The shape of the GZK feature is
strongly model-dependent: it is more flat in case of local
overdensity of the sources, and more steep in case of their local
deficit. It depends also on fluctuations in the distances between
sources inside the GZK sphere and on fluctuations of luminosities
of the sources there.
\begin{figure}[h]
\begin{center}
  \epsfig{figure=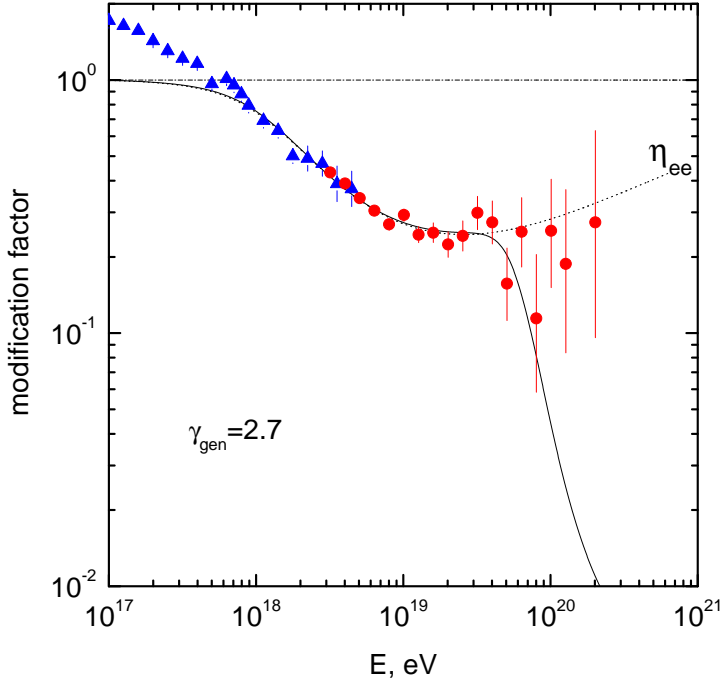,width=9.5cm}
\end{center}
%\vspace{-11mm}
\caption{
Predicted dip in comparison with the Akeno-AGASA data.
}
\label{mfactorA}
\end{figure}
The shape of the {\em dip} is fixed and has a specific form which
is difficult to imitate by other mechanisms. The protons in the
dip are collected from the large volume with the radius about
1000~Mpc and therefore the assumption of uniform distribution of
sources within this volume is well justified. In contrast to this
well predicted and specifically shaped feature, the cutoff, if
discovered, can be produced as the acceleration cutoff.
Since the shape of both GZK cutoff and
acceleration cutoff is model-dependent, it will be difficult to
argue in favour of any of them.  The problem of identification of
the dip depends on the accuracy of observational data, which
should confirm the specific (and well predicted) shape of this
feature. Do the present data have the needed accuracy?

The comparison of the calculated modification factor with that
obtained from the Akeno-AGASA data, using $\gamma_g=2.7$,
is given in Fig.~\ref{mfactorA}. It shows
the excellent agreement between predicted and observed
modification factors for the dip.

In Fig.~\ref{mfactorA} one observes that at $E < 1\times 10^{18}$~eV
the agreement between calculated and observed modification factors
becomes worse and at $E \leq 4\times 10^{17}$~eV the observational
modification factor becomes larger than 1. Since by definition
$\eta(E)\leq 1$, it signals about appearance of another
component of cosmic rays, which is most probably  galactic cosmic
rays. The condition $\eta
> 1$ means the dominance of the new (galactic) component, the
transition occurs at higher energy.
\begin{figure}[t]
\begin{center}
    \epsfig{figure=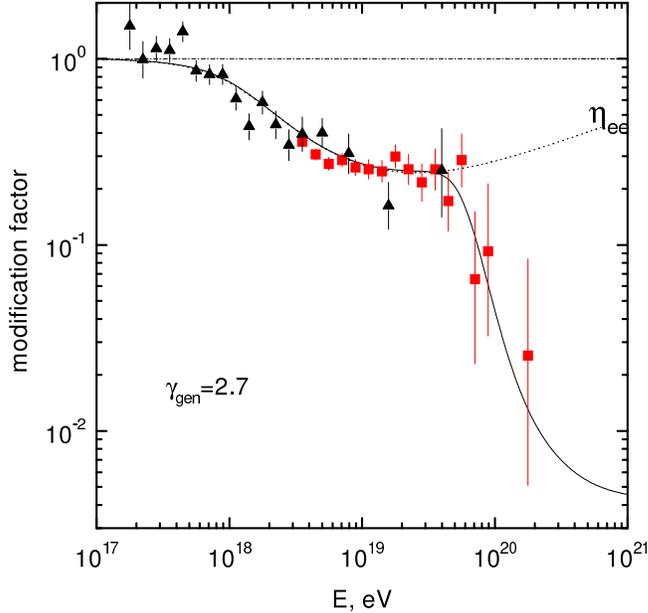,width=8.5cm}
\end{center}
\caption{
Predicted dip in comparison with the HiRes data.
}
\label{mfactorH}
\end{figure}
To calculate $\chi^2$ for the confirmation of the dip by
Akeno-AGASA data, we choose the energy interval between $1\times
10^{18}$~eV (which is somewhat arbitrary in our analysis) and
$4\times 10^{19}$~eV (the energy of intersection of $\eta_{ee}(E)$
and $\eta_{\rm tot}(E)$). In calculations we used the Gaussian
statistics for low-energy bins, and the Poisson statistics for the
high energy bins of AGASA. It results in $\chi^2=19.06$. The
number of Akeno-AGASA bins is 19. We use in calculations two free
parameters: $\gamma_g$ and the total normalization of spectrum. In
effect, the confirmation of the dip is characterised by
$\chi^2=19.06$ for d.o.f.=17, or $\chi^2$/d.o.f.=1.12, very close to
the ideal value 1.0 for the Poisson statistics.

In Fig.~\ref{mfactorH} the comparison of modification factor with
the HiRes data is shown. The agreement is also very good:
$\chi^2= 19.5$ for $d.o.f.= 19$ for the Poisson statistics.

The good agreement of the shape of the dip $\eta_{ee}(E)$ with
observations is a strong evidence for extragalactic protons
interacting with CMB. This evidence is confirmed by the HiRes data
on the mass composition (see Fig.~\ref{mass-comp}).

The dip is also present in case of diffusive propagation in magnetic
field \cite{AB1}.

\vspace*{-5mm}
\section{Extragalactic iron nuclei as UHECR primaries}
\label{iron_section} \vspace*{-5mm}
Does modification factor for
iron nuclei differ from the proton dip?
\begin{figure}[ht]
\begin{center}
\epsfig{figure=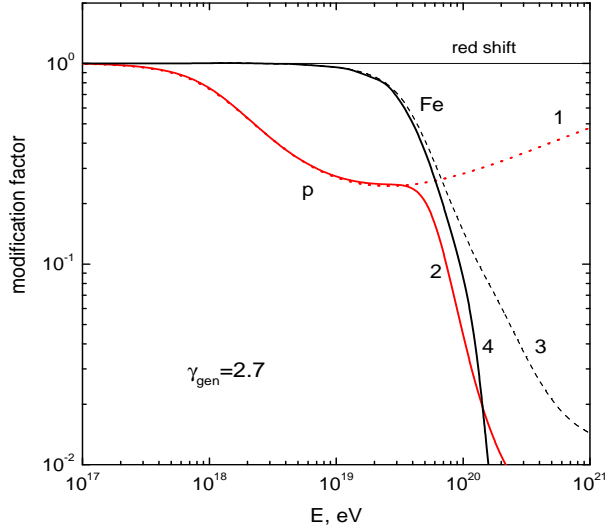,height=7cm,width=8cm}
\end{center}
%\vspace{-10mm}
\caption{
Modification factor for iron nuclei in comparison with that
for protons. Curve $\eta=1$ corresponds to adiabatic energy losses.
Proton modification factors are given given by curve 1 (adiabatic and
pair production energy losses) and by curve 2 (total energy losses).
Iron modification factors are given by curve 3 (adiabatic and
pair production energy losses) and by curve 4 (with photodissociation
included).
}
\label{iron}
\end{figure}

We calculated the modification factor for iron nuclei, assuming that 
that Fe nuclei are the heaviest ones accelerated in the sources, 
and considering the propagation of Fe nuclei with energy losses taken
into account. The resulting flux is given only for primary iron nuclei,
without secondary nuclei produced during propagation (more details
will be presented in \cite{ABG}). 

The energy losses for Fe are dominated by adiabatic energy losses
up to $4.5\times 10^{19}$~eV, from where on $e^+e^-$ energy losses
dominate. For energies $E \geq 1.7 \times 10^{20}$~eV
photodissociation becomes the main source of energy losses
\cite{Stecker,ABG}. According to this dependence the energy
spectrum of iron nuclei is $\propto E^{-\gamma_g}$ up to $E_1 \sim
1\times 10^{19}$~eV, where the first steepening begins. The
second steepening, caused by iron-nuclei destruction, occurs at
energy $E_2 \sim 1\times 10^{20}$~eV. For lighter nuclei the
steepening (cutoff) starts at lower energies. Therefore the cutoff
of the nuclei spectra occurs approximately at the same energy as
the GZK cutoff, though the physical reason for these two cutoffs is
different: while the latter (GZK) is caused by starting of photopion
production, the former (nuclei) - by transition from adiabatic to
pair-production energy losses \cite{BGZ}.

In Fig.~\ref{iron} the modification factors for iron nuclei are
shown as function of energy in comparison with modification
factors for protons.  Comparison with Fig.~\ref{mfactorA} clearly shows
that even small admixture of iron nuclei in the primary extragalactic
flux upsets
the good agreement of the proton dip with observational data.
\vspace*{-5mm}
\section{Discussion and conclusions}
\label{discussion}
\vspace*{-5mm}
There are three signatures of UHE protons propagating through CMB:
GZK cutoff, bump and dip.

The energy shape of the GZK feature is model dependent. The local
excess of sources makes it flatter, and the deficit - steeper. The
shape is affected by fluctuations of source luminosities and
distances between the sources. The cutoff, if discovered, can be
produced as the acceleration cutoff (steepening below the maximum
energy of acceleration). Since the shape of both, the GZK cutoff
and acceleration cutoff, is model-dependent, it will be difficult
to argue in favour of any of them, in case a cutoff is discovered.

The {\em bump} is produced by pile-up protons, which are loosing
energy in photopion interactions and are accumulated at low energy,
where the photopion energy losses become
equal to that due to pair-production. Such bump is distinctly seen in
calculation of spectrum from a single remote source. In the diffuse
spectrum, since the individual peaks located at different energies,
a flat spectrum feature is produced.

The {\em dip} is the most remarkable feature of interaction with
CMB. The protons in this energy region are collected from the
distances $\sim 1000$~Mpc, with each radial interval $dr$
providing the equal flux. All density irregularities and all fluctuations
are averaged at this distance, and assumption of uniform distribution
of sources with average distances between sources and average
luminosities becomes quite reliable. The dip is confirmed by
Akeno-AGASA and HiRes data with the great accuracy (see
Figs \ref{mfactorA} and  \ref{mfactorH}). As one can see from
Fig.~\ref{iron},
presence of even small fraction of extragalactic heavy nuclei in the
primary flux upsets this agreement.\\*[2mm]
\noindent
{\em We interpret the excellent agreement of the calculated dip with the
observations as an independent evidence that observed primaries at energy
$1\times 10^{18} - 4\times 10^{19}$~eV are extragalactic protons.
This evidence is the complementary one to the direct measurements
(now contradictive) of chemical composition.}\\*[2mm]
\noindent
At energy
$E< 4\times 10^{17}$~eV the modification factor from Akeno data
exceeds 1, and it signals about dominance of another cosmic ray
component, most
probably the galactic one. It agrees with transition from galactic to
extragalactic component at the second knee $E \sim 1\times 10^{18}$~eV.
This conclusion is confirmed by the recent HiRes data  on mass composition
(see Fig.~\ref{mass-comp}) and indirectly by the KASCADE data
(see \cite{BGH} for the detailed analysis).

\begin{figure}[h]
\begin{center}
\epsfig{figure=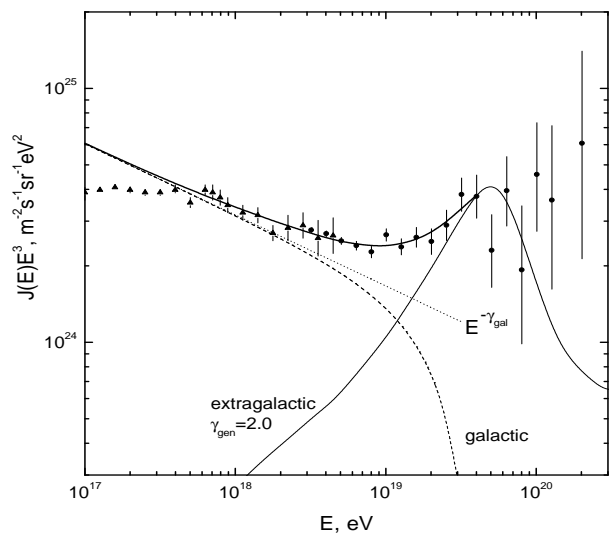,height=7cm,width=8cm}
\end{center}
%\vspace{-10mm}
\caption{
The ad hoc model for explanation of the dip. The distorted
power-law spectrum (\ref{gal}) is shown by curve ``galactic''.
The extragalactic spectrum with $\gamma_g=2$ is normalized by the data
in the interval $(4 - 6)\times 10^{19}$~eV. The parameters of the
galactic spectrum are found to provide the best fit to the data by the total
spectrum (galactic + extragalactic), shown by the thick curve. It gives
$\gamma_{\rm gal}=3.28$.
}
\label{adhoc}
\end{figure}

Are there alternative explanations of the dip? The conservative one
(see e.g. \cite{ankle})
is known since early 80s, when the spectrum feature, {\em ankle}, was
discovered in the Haverah Park data \cite{HP80} at $E \sim 1\times 10^{19}$~eV.
This feature was interpreted as transition from galactic to
extragalactic cosmic rays (in contrast to the calculations above where
the ankle  naturally appears as a part of the dip). The hypothesis of the
transition at the ankle can be described phenomenologically as follows:
At energy below $1\times 10^{19}$~eV
the cosmic ray flux is galactic and above - extragalactic. The galactic
spectrum can be taken basically as
power-law $\propto E^{- \gamma_{\rm gal}}$,
but agreement with observations
needs steepening at $E \geq E_g$, described by some steepening parameter.
In effect one can use the parametrization:
\begin{equation}
I_{\rm gal}(E)=K_{\rm gal}E^{-\gamma_{\rm gal}}
\left [1- a\exp(b\log E/10^{19}~{\rm eV}) \right ].
\label{gal}
\end{equation}
The extragalactic generation  spectrum is assumed to be power-law
with index $\gamma_g$. Together with two constants of the
normalization for both fluxes, one has as minimum six free
parameters to fit the observed spectrum.  We found the best fit
shown in Fig.~\ref{adhoc}. It is characterized by $\chi^2=9.1$ for
19 energy bins and 6 free parameters, i.e. by  $\chi^2/d.o.f.=
0.7$ for d.o.f.=13. The value $\chi^2/d.o.f. <1$ for the Poisson
statistics signals for the large number of the free parameters and
for very good formal fit to the experimental data. The problem of
this ad hoc model is whether there is a physical model for
propagation of galactic cosmic rays, which results in spectrum
given by Eq.(\ref{gal}). It is just {\em assumed} that the
spectrum at $E < 1\times 10^{19}$~eV is the same as observed,
while the dip model {\em predicts} this spectrum in excellent
agreement with observations. An intermediate case is given in
\cite{Dermer} where the dip is mostly described by extragalactic
protons interacting with CMB with a small correction given by
galactic cosmic rays only at the low-energy, $E\approx 1\times
10^{18}$~eV, part of the dip
(see Fig. 13 in \cite{Dermer}).\\
How does extragalactic magnetic field affect the discussed spectra
features?

The influence of magnetic field on spectrum depends on
the separation of the sources $d$. There is a statement which has a
status of the theorem \cite{AB}:\\*[2mm]
For uniform distribution of sources
with separation much less than characteristic lengths of
propagation, such as energy attenuation length $l_{\rm att}$
and the diffusion length $l_{\rm diff}$, the diffuse spectrum of
UHECR has an universal (standard) form independent of mode of
propagation.

For the realistic intergalactic magnetic fields the spectrum is 
universal in energy interval $1\times 10^{18} - 8\times 10^{19}$~eV 
\cite{AB1}. Note, however, that
generation spectrum is defined in \cite{AB} as one outside the
source. In this work we implicitly assume that the sources are
transparent for UHE protons and thus the generation spectrum is the
same as the acceleration spectrum.

The most probable astrophysical sources of UHECR are AGN. They can
accelerate particles to $E_{\rm max} \sim 1\times 10^{21}$~eV and
provide the needed emissivity of UHECR
${\mathcal L}_0 \sim 3\times 10^{46}$~erg/Mpc$^3$yr. The correlation of
UHE particles with directions to special type of AGN, Bl Lacs, is
found in analysis of work \cite{TT}. AGN as UHECR sources in case of
quasi(rectilinear) propagation of protons explain most naturally
the small-scale anisotropy \cite{small-scale}.

The UHECR from AGN have a problem with
superGZK particles with energies $E> 1\times 10^{20}$~eV: (i) another
component is needed for explanation of the AGASA excess, and (ii)
no sources are observed in AGASA and other arrays in direction of
superGZK particles. These problems probably imply the new physics,
such as UHECR from superheavy dark matter, new signal carrier, like e.g.
new light stable hadron, strongly interacting neutrino, and Lorentz
invariance violation. For the last case it is interesting to note that
if Lorentz invariance is weakly broken for $e^+e^-$ production, but
strongly for pion production, then the modification factor is given by
the curve $\eta_{ee}$ in Fig.~\ref{mfactorA}. This prediction agrees
well with the Akeno-AGASA spectrum.

\vspace*{-7mm}
\section*{Acknowledgements}
\vspace*{-7mm}
We thank transnational access to research
infrastructures (TARI) program through the LNGS TARI grant
contract HPRI-CT-2001-00149. The work of S.I.G. was partly
supported by grants RFBR 03-02-1643a and RFBR 04-0216757a.

\end{document}